\documentclass[journal=nanoletters,manuscript=article]{achemso}
\setkeys{acs}{articletitle=true}

\usepackage{chemformula} % Formula subscripts using \ch{}
\usepackage[T1]{fontenc} % Use modern font encodings
\usepackage{siunitx}

\author{Andr\'as Kov\'acs}
\affiliation{Ernst Ruska-Centre for Microscopy and Spectroscopy with Electrons, Peter Gr\"unberg Institute, Forschungszentrum J\"ulich GmbH, 52425 J\"ulich, Germany}
\email{a.kovacs@fz-juelich.de}
\author{Laura H. Lewis} 
\affiliation
{Department of Chemical Engineering, Northeastern University, Boston, MA 02115, USA}
\affiliation{Department of Mechanical and Industrial Engineering, Northeastern University, Boston, MA 02115, USA}
\author{Dhanalaksmi Palanisamy}
\affiliation{Max-Planck-Institut f\"ur Eisenforschung, 40237 D\"usseldorf, Germany}
\author{Thibaud Denneulin}
\affiliation{Ernst Ruska-Centre for Microscopy and Spectroscopy with Electrons, Peter Gr\"unberg Institute, Forschungszentrum J\"ulich GmbH, 52425 J\"ulich, Germany}
\author{Alexander Schwedt}
\affiliation{Central Facility for Electron Microscopy, RWTH Aachen University, 52074 Aachen, Germany}
\author{Edward R.D. Scott}
\affiliation{Hawaii Institute of Geophysics and Planetology, University of Hawaii, Honolulu, HI 96822, USA}
\author{Baptiste Gault}
\affiliation{Max-Planck-Institut f\"ur Eisenforschung, 40237 D\"usseldorf, Germany}
\affiliation{Department of Materials, Royal School of Mines, Imperial College London, UK}
\author{Dierk Raabe}
\affiliation{Max-Planck-Institut f\"ur Eisenforschung, 40237 D\"usseldorf, Germany}
\author{Rafal E. Dunin-Borkowski}
\affiliation{Ernst Ruska-Centre for Microscopy and Spectroscopy with Electrons, Peter Gr\"unberg Institute, Forschungszentrum J\"ulich GmbH, 52425 J\"ulich, Germany}
\author{Michalis Charilaou}
\affiliation{Department of Physics, University of Louisiana at Lafayette, Lafayette, Louisiana 70504, USA}
\email{michalis.charilaou@louisiana.edu}

\title
  {Discovery and implications of hidden atomic-scale structure in a metallic meteorite}

\begin{document}

\begin{tocentry}

Some journals require a graphical entry for the Table of Contents.
This should be laid out ``print ready'' so that the sizing of the
text is correct.

Inside the \texttt{tocentry} environment, the font used is Helvetica
8\,pt, as required by \emph{Journal of the American Chemical
Society}.

The surrounding frame is 9\,cm by 3.5\,cm, which is the maximum
permitted for  \emph{Journal of the American Chemical Society}
graphical table of content entries. The box will not resize if the
content is too big: instead it will overflow the edge of the box.

This box and the associated title will always be printed on a
separate page at the end of the document.

\end{tocentry}

\begin{abstract}
 Iron and its alloys have made modern civilization possible, with metallic meteorites providing one of the human's earliest sources of usable iron as well as providing a window into our solar system's billion-year history. Here highest-resolution tools reveal the existence of a previously hidden FeNi nanophase within the extremely slowly cooled metallic meteorite NWA 6259. This new nanophase exists alongside Ni-poor and Ni-rich nanoprecipitates within a matrix of tetrataenite, the uniaxial, chemically ordered form of FeNi. The ferromagnetic nature of the nanoprecipitates combined with the antiferromagnetic character of the FeNi nanophases give rise to a complex magnetic state that evolves dramatically with temperature. These observations extend and possibly alter our understanding of celestial metallurgy, provide new knowledge concerning the archetypal Fe-Ni phase diagram and supply new information for the development of new types of sustainable, technologically critical high-energy magnets.
\end{abstract}

%\section{Introduction}
The study of ferrous meteorites informs our understanding of the solar system as well as of terrestrial metallurgy. These bodies consisting primarily of iron and nickel, are remnants of protoplanetary cores that formed during the early solar system \cite{krot2008,goldstein2009,weiss2013,Scott2020} and are thought to have produced magnetic fields in a similar manner to Earth's geodynamo \cite{weiss2013}. Although the original location of iron meteorites is thought to be the asteroid belt, i.e., between the orbits of Mars and Jupiter, isotopic measurements suggest that some meteorites originated beyond Jupiter \cite{kruijer2017}, while others came from the Earth-forming region in the interior of the solar system. Therefore, the study of metallic meteorites, which provide the oldest thermal and magnetic record of the early solar system, can provide a deep understanding of what may have been the precursor of Earth itself. 
From a materials science perspective, meteorites provide almost ideal environments for atomic arrangements to approach thermodynamic equilibrium during cooling over billions of years. Such conditions can  permit the formation of tetrataenite (designation L$1_0$, AuCu-I prototype structure) which is extremely difficult to synthesize in the laboratory \cite{poirier2015,lewis2014}. Tetrataenite's alternating layers of Fe and Ni atoms are stacked parallel to the tetragonal \textit{c} axis to form a superlattice that donates impressive technical magnetic properties \cite{Lewis2020}. Tetrataenite is not documented in conventional Fe-Ni binary phase diagram \cite{Swartzendruber1991} but  may be found in meteoritical phase diagrams \cite{Scott2020,yang1996,howald2003,scorzelli2015} containing a complex set of ferromagnetic phases \cite{Scott2020,scott1997,yang2007} that are, by convention, designated by their Ni content. The L$1_0$ phase of FeNi forms during cooling from disordered face-centered cubic (fcc, designation A1) Ni-rich taenite. Other meteoritic phases include kamacite, the Ni-poor body-centered cubic (bcc, designation A2) alloy that contains a maximum of 5 at.\% Ni \cite{owen1969,clarke1978,yang2005}, and awaruite, an intermetallic Ni$_3$Fe-type compound with L$1_2$-type structure \cite{howald2003,yang1997}. These Fe-Ni phases and their crystallographic information are summarized in Supporting Information. The kinetics of phase transformations in the Fe-Ni system are acknowledged to be extremely slow \cite{scorzelli2015} as a result of the sluggish interdiffusion of Fe and Ni, which is likely influenced by magnetic long-range order \cite{goldstein2009}. Details of the phase assemblage in a meteorite determine its internal magnetic field \cite{Skomski2013,harrison2018}, which impacts the interpretation of its thermal history \cite{goldstein2009,yang1997}.

\subsection{A clandestine meteoritic microstructure}
\begin{figure*}[t]
	\centering
	\includegraphics[width=\linewidth]{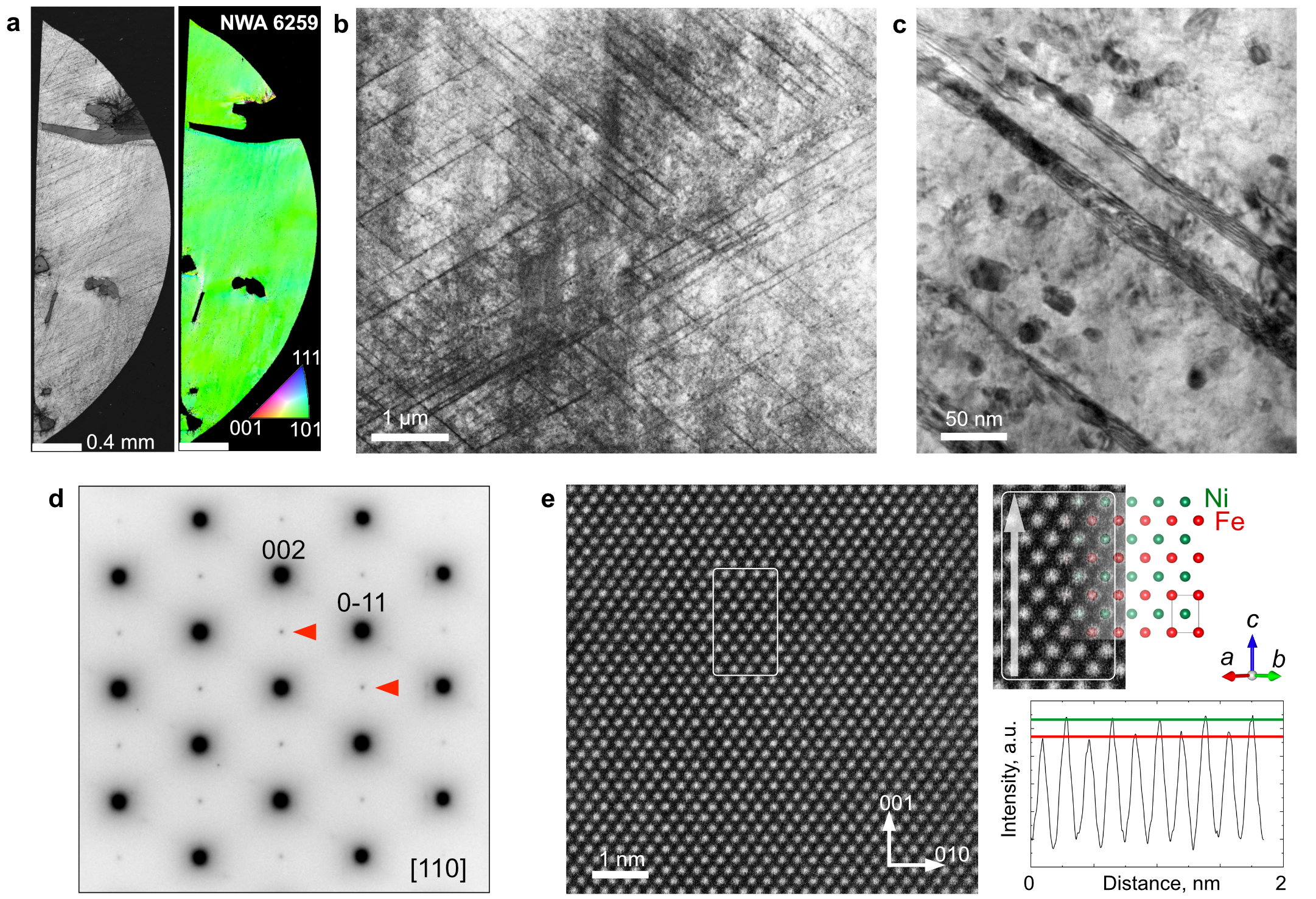}
	\caption{Microstructure of meteorite NWA 6259. 
		(a) Electron backscatter diffraction image quality map recorded from the fragment used for electron microscopy. The dark grey inclusions contain sulfur and phosphorus. The inverse pole figure map with respect to the sample normal direction on the right shows that the crystal is one grain.
		(b) Bright-field TEM image showing lamellar microstructure.
		(c) Magnified bright-field TEM image revealing precipitates adjacent to lamellae in the Fe-Ni matrix. 
		(d) Electron diffraction pattern recorded from the area shown in (b), consistent with an ordered tetragonal L$1_0$ structure. The viewing direction is [100]. Red triangles mark superlattice reflections. (e) HAADF STEM image of the tetrataenite L$1_0$ FeNi matrix. The enlarged region and schematic diagram on the right shows a primitive unit cell of the tetragonal phase. Intensity variations in the line profile, which was obtained from the marked region, are associated with differences in atomic number Z between Fe (Z = 26) and Ni (Z = 28). The detector semi-angle used was 69 mrad.}
	\label{fig:tem}
\end{figure*}

In this work, the investigation was focused on the NWA 6259 meteorite which consists of a very large multi-variant region of tetrataenite \cite{poirier2015} and possesses the second highest Ni content ($\sim$43 at.\%) of all reported meteorites. The structure of the NWA 6259 specimen is shown on different length scales in Figure \ref{fig:tem}. Details of sample preparation and characterization techniques are provided in the Supporting Information (Fig.S3). A sample for study was removed from the central region of the meteorite specimen (Figure \ref{fig:tem}a) and was determined to possess an approximate mesoscopic composition of Fe 57 at.\% and Ni 43 at.\%; with Co ($\sim$3 at.\%) and a minor enrichment in Cu (Fig. S4); the dark inclusions observed in Figure \ref{fig:tem}a contain sulphur and phosphorus. A crystallographic orientation map derived from electron backscatter diffraction (EBSD) data (Figure \ref{fig:tem}a) reveals that, within the resolution limit of the technique, this region can be considered as a single crystal. This orientation map guided the preparation of crystallographically-defined electron-transparent specimens (Figure \ref{fig:tem}b, Fig. S3) for higher resolution studies. The specimen matrix contains a network of precipitates and lamellar inclusions (Figures \ref{fig:tem}b,c) and is verified to possess tetragonal symmetry with superlattice diffraction reflections (Figure \ref{fig:tem}d) that signal the long-range chemical order of L$1_0$ FeNi, tetrataenite. The high degree of chemical order of the tetrataenite matrix is confirmed by the small but finite intensity difference attributed to alternate scattering of Fe and Ni atom columns detected by high-angle annular dark field (HAADF) scanning transmission electron microscopy (TEM) (Figure \ref{fig:tem}e). 

\begin{figure}[t]
	\centering
	\includegraphics[width=0.6\linewidth]{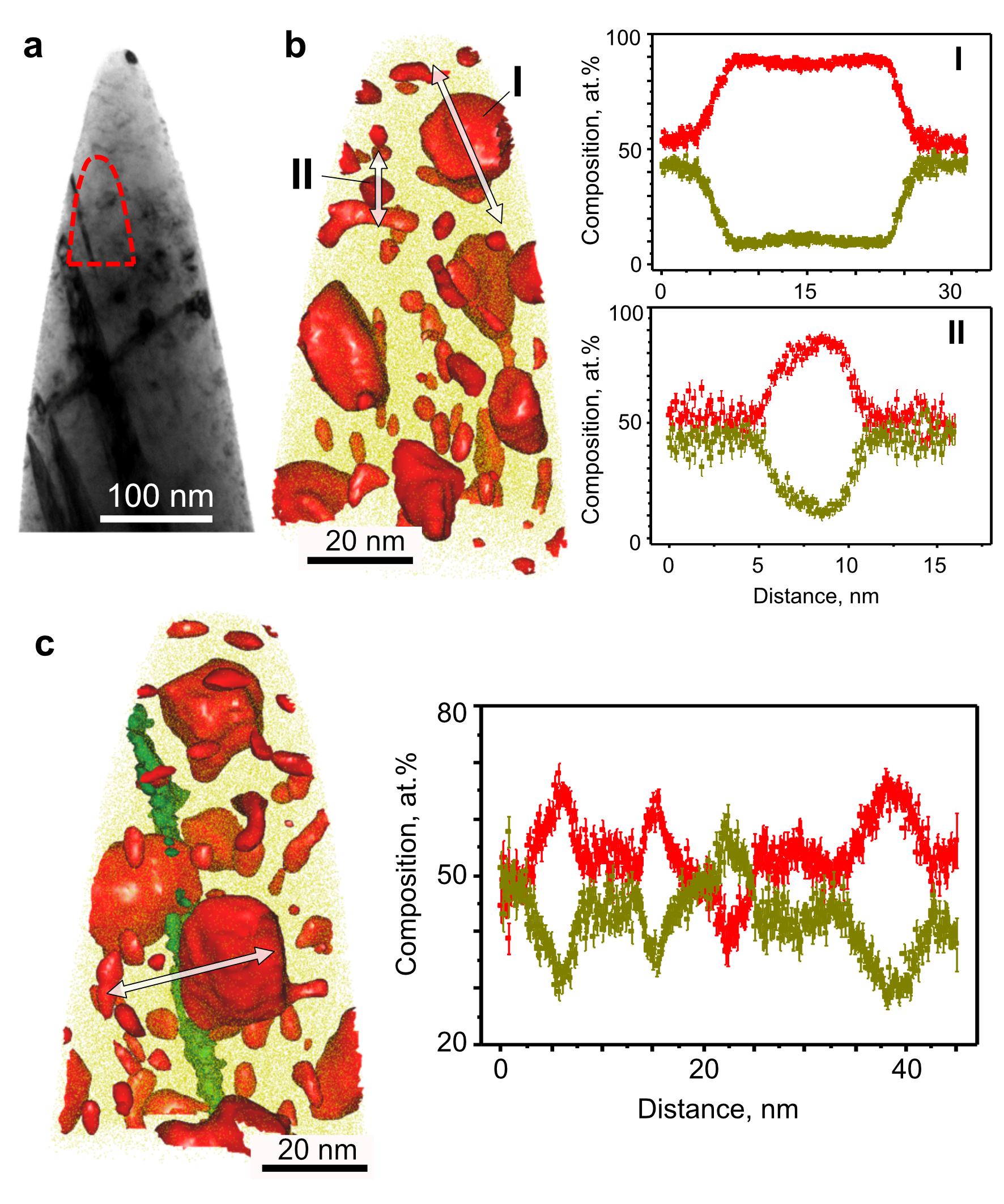}
	\caption{Fe-Ni phase decomposition in tetrataenite matrix. (a) Bright-field TEM image of a needle-shaped specimen prepared for atom probe tomography reconstruction. The marked region (red cone) is reconstructed and analyzed in (b). (b) Reconstruction showing Ni-poor (Fe-rich) regions (red) delineated by 26 at.\% Ni iso-concentration surfaces. The corresponding elemental concentration profiles (I and II) across the particles marked in (b) (Fe: red/ Ni: green), showing an Fe composition close to 90 at.\%. 
		(c) Reconstruction of the precipitates and lamella delineated by a 50 at.\% Ni iso-concentration surface. Based on the Fe-Ni phase diagram, the lamella is inferred to be awaruite, fcc FeNi$_2$. Corresponding composition profile along the line marked in (c) showing Fe and Ni enrichment in the tetrataenite matrix. }
	\label{fig:apt}
\end{figure}

The structure and composition of small (Figure \ref{fig:tem}c) crystalline precipitates within the L$1_0$ matrix were examined at \AA ngstrom-level resolution using correlative electron microscopy and 3-dimensional (3D) atom probe tomography (APT) performed on the needle-shaped specimen (Figure \ref{fig:apt}a) prepared using focused ion beam (FIB) milling. Theses results confirm the tetrataenite composition itself as 45 at.\% Ni: 55 at.\%; within the matrix, iso-composition surfaces superimposed onto the reconstructed tomographic 3D point cloud reveal that regions richer than 26 at.\% Ni (Figure \ref{fig:apt}b) contain a dense distribution of Ni-poor ($\sim$90 at.\% Fe) precipitates with a bimodal distribution of coarse (28$\pm$6 nm) and ultrafine (2.0$\pm$0.5 nm) average diameters at approximately 15000 precipitates per cubic micrometer (Figures \ref{fig:stem}). The 50 at.\% iso-composition level reveals Ni-rich lamellae of composition of $\sim$66 at.\% Ni, close to that of the ideal composition of the awaruite \cite{howald2003,Wakelin_1953}, Figure \ref{fig:apt}c. A combined tomographic reconstruction in Figure \ref{fig:apt}c shows the overall nanostructure, together with corresponding quantitative elemental composition scans. 

\subsection{Nanostructure, strain and a new Fe - Ni phase}

\begin{figure}[t]
	\centering
	\includegraphics[width=\linewidth]{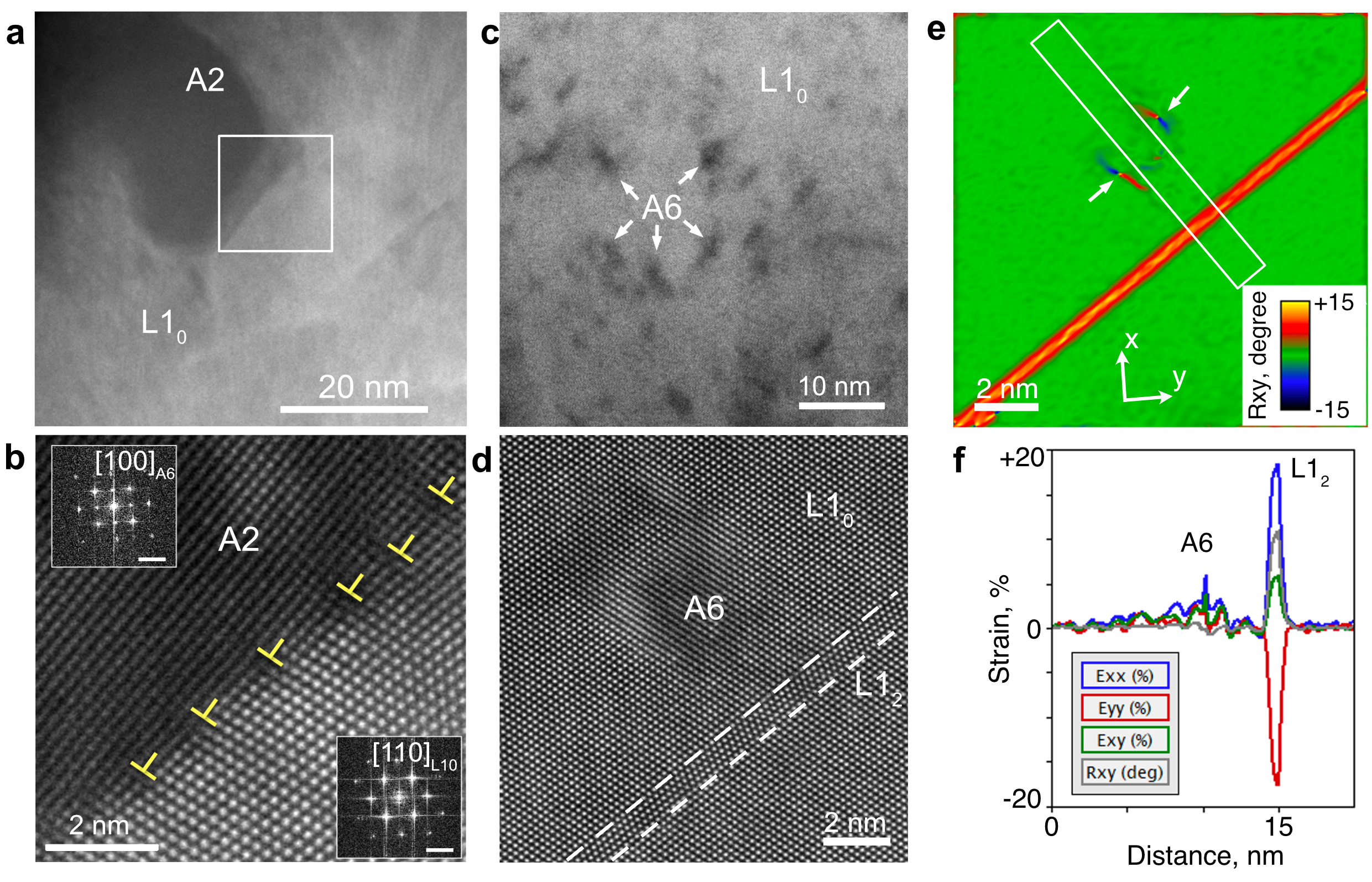}
	\caption{Structure and strain analyses. (a) Overview HAADF STEM image of a bcc A2 Fe-Ni precipitate (25 x 35 nm), shown alongside (b) an atomic-resolution HAADF STEM image of the A2/L$1_0$ interface, which is decorated by misfit dislocations every 1-2 nm. The inset Fourier transforms confirm the bcc structure of the precipitate. (c) Overview bright-field STEM image of ultrafine (<5 nm) Ni-poor A6 Fe-Ni precipitates (dark). (d) Atomic-resolution HAADF STEM image of an A6 precipitate next to a 3-monolayers-thick Ni-rich L$1_2$ lamella. (e) Strain rotation map of the A6 and L$1_2$ lamella shown in (d). Arrows mark misfit dislocations at the precipitate boundary. (f) Strain and shear in the region marked in (e) across the A6 precipitate and L$1_2$ lamella. }
	\label{fig:stem}
\end{figure}

A fascinating aspect of the meteorite nanostructure is the role that strain plays in the crystallographic features of two types of Ni-poor precipitates embedded within the L$1_0$-type matrix. A representative coarse Ni-poor precipitate, embedded incoherently in the matrix, is delineated by regularly spaced (1 - 2 nm) misfit dislocations at the precipitate-matrix interface ([100]$_\mathrm{A2}$(010)$_\mathrm{A2}$~||~[110]$_{\mathrm{L1_0}}$(001)$_{\mathrm{L1_0}}$ orientational relationship) and is confirmed to adopt the bcc (A2-type) structure (Figure \ref{fig:stem}a,b). In contrast, the ultrafine (1 - 2 nm) Ni-poor precipitates, Fig. \ref{fig:stem}c,d, coherently embedded in the matrix, have the same crystal symmetry as the surrounding L$1_0$ matrix but possess a \textit{chemically disordered} face-centered tetragonal (A6-type) crystal structure with unit cell parameters similar to tetrataenite. These ultrafine precipitates are lattice-matched to the tetrataenite matrix but possess a composition of 90 at.\% Fe. To the best of our knowledge, this is the first report of a tetragonal Ni-poor phase in the Fe-Ni system, although recently the synthesis of tetragonal, nominally equiatomic FeNi has been confirmed \cite{Lewis2020,lewis2016}.

A relatively big, 4-nm-diameter Ni-poor precipitate, adjacent to a Ni-rich lamella, Fig. \ref{fig:stem}d, is characterized by a strain field as revealed by geometric phase analysis based on Fourier transformation of a high-resolution STEM image as shown in Figure \ref{fig:stem}e,f. This region, which contains two dislocations and a corresponding strain at the phase boundary (Figure \ref{fig:stem}e), is consistent with the interpretation of nanoscale decomposition of the metastable tetrataenite phase through precipitation of Ni-poor phases with either cubic A2 (coarse kamacite precipitates) or tetragonal A6 (utrafine precipitates) crystal structures and a lamellar Ni-rich L$1_2$-type phase. While the two types of Ni-poor precipitates are nearly isotropic in shape and are distributed evenly in the matrix, the Ni-rich awaruite precipitates follow distinct crystallographic directions in the matrix, suggesting that Ni migrated along diffusion-favorable directions to form the lamellae, leaving behind Ni-poor pockets. 

The tetragonal A6-structured FeNi phase in this meteorite divulges a fascinating new aspect of the Fe - Ni system. The close relationship between various cubic symmetries was formalized long ago as the Bain distortion \cite{bain1924}, in which a bcc lattice can be obtained from an fcc lattice by a compression parallel to the c axis and an expansion along an a axis to form a body-centered tetragonal lattice \cite{bowles1972}. The A6 structure adopted by the ultrafine Ni-poor precipitates is likely stabilized through largely coherent bonding to the parent tetragonal L$1_0$ phase. These ultrafine A6 precipitates can be considered as Guinier-Preston (G.P.) zones, which are manifestations of an initial stage of precipitation during solid-state phase decomposition \cite{hill1973}. G.P. zones typically possess an intermediate crystal structure and composition that are different from those of both the thermodynamically stable phase and the host phase.

\section*{Effects of phase diversities on magnetic properties}

\begin{figure}[b]
	\centering
	\includegraphics[width=\linewidth]{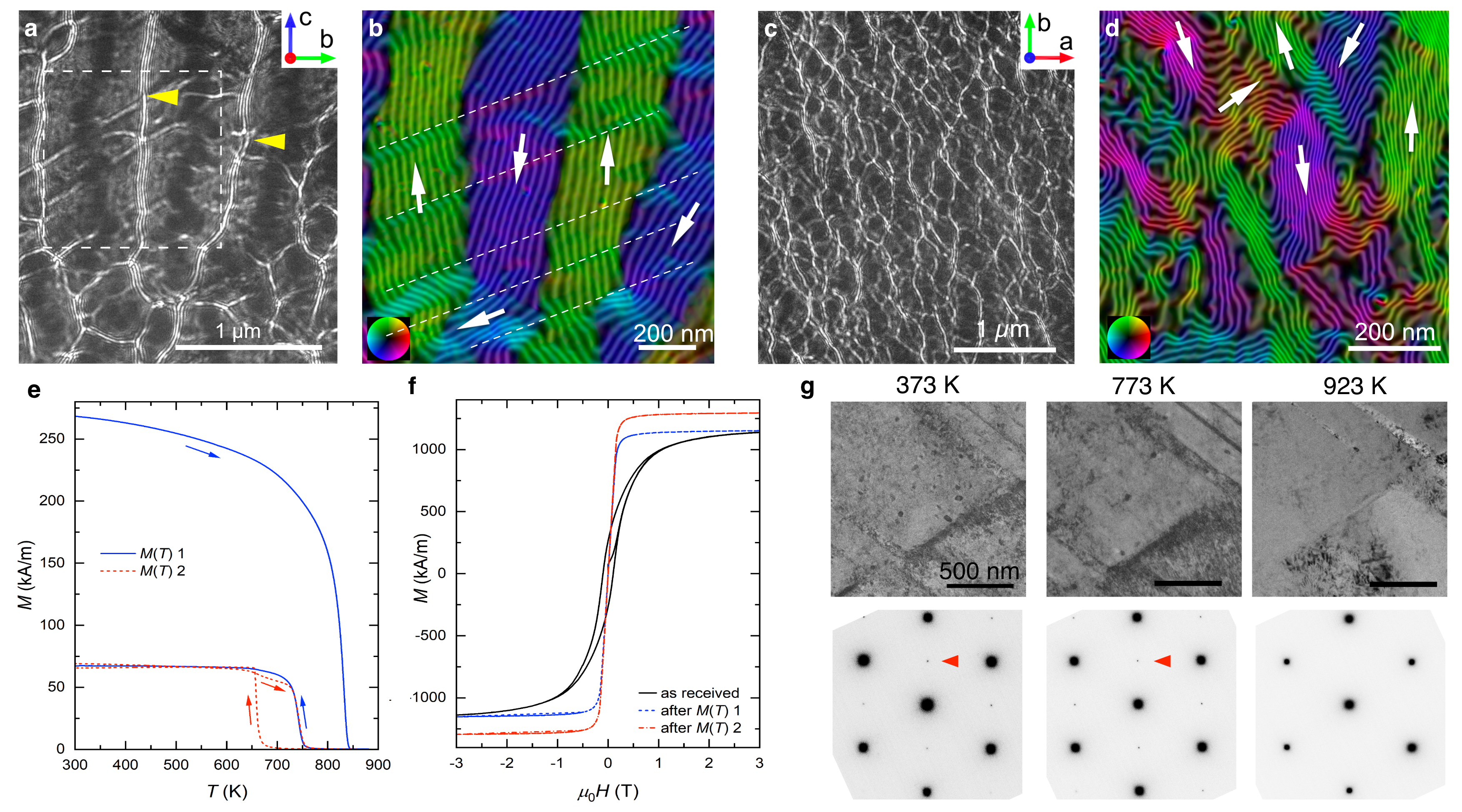}
	\caption{Magnetic properties of NWA 6259. (a) Fresnel defocus image of a specimen in which the magnetic easy axis ([001] of L$1_0$, \textit{c} axis) is in-plane. Bands of black and white contrast arise from the presence of magnetic domain walls. Variations in contrast marked with yellow triangles suggest local changes in wall inclination. (b) Magnetic induction map measured using off-axis electron holography from the dashed region in (a), showing \ang{180} magnetic domain walls. Dashed lines mark the locations of Ni-rich lamella precipitates. Colors and arrows indicate the magnetic field direction. Contour spacing: 2$\pi$ radians. (c) Fresnel defocus image of a specimen in which the magnetic easy axis is out-of-plane. Defocus: 0.5 mm. (d) Magnetic induction map showing a complex arrangement of magnetic domains. Contour spacing: 2$\pi$ radians. (e) Thermo-magnetic curves M(T) 1 and M(T) 2. The ferromagnetic transition temperature is 830 K. (f) Magnetization hysteresis curves measured in the ``as received'' condition and after annealing cycles (M(T) 1, 2). (g) Precipitate dissolution and chemical disordering observed in bright-field TEM images and SAED patterns recorded at 373, 773 and 923 K. Red triangles mark 001 reflections of the ordered L$1_0$ structure. }
	\label{fig:magn}
\end{figure}

The presence and diversity of these Fe-Ni phases impacts both the micromagnetic and bulk magnetic states of the material, and consequently influences how magnetometry is used to interpret meteoritic history as well as to evaluate tetrataenite's potential as a technological material. The results reported here confirm that the NWA 6259 meteorite, and therefore likely other stony, stony-iron and iron meteorites, can be regarded as magnetic nanocomposites with strong interphase magnetic coupling. Magnetic configurations in nanocomposites have been studied extensively as novel exchanged-coupled permanent magnets \cite{Skomski2013,skomski2003}, and it is known that extrinsic, or technical, magnetic properties such as coercivity and remanence depend on the volume fractions of the phases, the diameters of precipitates and the degree of exchange coupling at interfaces. In order to investigate these aspects, the magnetic configuration of the NWA 6259 meteorite was studied using Lorentz TEM and off-axis electron holography \cite{Kovacs2018} (Figures \ref{fig:magn}a-d) applied to the same samples that were investigated using microstructural characterization (Supporting Information, Fig. S5). Magnetic imaging in the remanent state was conducted in specimens prepared with the magnetic easy axis of the L$1_0$ FeNi phase oriented both in-plane (Figures \ref{fig:magn}a,b) and out-of-plane (Figures \ref{fig:magn}c,d). These images reveal a high density of  \ang{180} or \ang{90} magnetic domains, with sizes ranging from 100 to 500 nm. Quantitative magnetic induction maps (Figs \ref{fig:magn}b,d) indicate that the \ang{180} magnetic domain walls are almost parallel to the magnetic easy axis, as expected for a uniaxial system \cite{kittel1949}. The magnetic domain walls are distorted in the vicinity of the Ni-rich lamellae, marked by dashed lines. This distortion is attributed to the difference in magnetocrystalline anisotropy energy of the L$1_0$ and L$1_2$ Fe-Ni phases. The Ni-poor A2 nano- and A6 ultrafine precipitates are not observed to affect the overall magnetic domain configuration in the studied samples. Nonetheless, all phases impact the magnetic state, and the nature of the A6-type (tetragonal) precipitates is of particular interest. Computational \cite{pinski1986,moruzzi1989} and experimental \cite{tsunoda1988a,tsunoda1988b} investigations of fcc-type iron indicate an anti-ferromagnetic \cite{lavrentiev2014,abrikosov2007,xiong2011,sjostedt2002} (AFM) ground state that is typically not accessible because the A2 (bcc) to A1 (fcc) phase transition in iron occurs above its Curie temperature (\textit{i.e.}, the temperature below which it is ferromagnetic). An atomistic simulation of a non-collinear configuration of atomic moments that leads to zero net magnetization (Fig. S6). As AFM ordering breaks cubic symmetry, antiferromagnetism is intimately linked to the presence of a structural distortion \cite{marsman2002,massalski2009,Dunin-Borkowski1999}. Thus, the A6-type tetragonal Fe(Ni) phase stabilized in the NWA 6259 meteorite is anticipated to exhibit antiferromagnetism. This hypothesis was investigated with bulk thermomagnetic measurements conducted in low magnetic field on a single sample of the as-received NWA 6259 meteorite. Two consecutive heating and cooling cycles in the temperature range 300 K $\leq$ T $\leq$ 900 K (Figure \ref{fig:magn}e) were performed, with full magnetic hysteresis loop measured at room temperature before and after each thermal excursion (Figure \ref{fig:magn}f). The first heating branch confirmed the reported tetrataenite kinetic Curie temperature $T_{C1}$ $\sim$ 830 K \cite{poirier2015,scorzelli2015}. Upon the first cooling, the magnetization remained close to zero until an apparent second Curie temperature of $T_{C2}$ $\sim$ 740 K where it rose to a value of 65 kA/m that was maintained down to room temperature (Figure \ref{fig:magn}f). The corresponding hysteresis loop returned a room temperature saturation magnetization of 1150 kA/m, same as that of the as-received state, but with a vanishingly small coercivity much decreased from the as-received value of 0.1 T, as expected for the chemically disordered FeNi phase. The low-field magnetization of the second heating cycle dipped slightly at $\sim$ 660 K and then fell abruptly at $T_{C2}$ $\sim$ 740 K. Upon the final cooling from 900 K, the magnetization again remained at an extremely low value down to a new magnetic transition at $T_{C3}$ $\sim$ 660 K to rise again to 65 kA/m. Most strikingly, the final hysteresis loop indicated a 14 \% increase of the room temperature saturation magnetization to $\sim$ 1260 kA/m (Figure \ref{fig:magn}f) with coercivity still at  nearly zero. These results motivated an \textit{in situ} annealing study in the TEM (Figure \ref{fig:magn}g), which indicates dissolution of the noted precipitates and concurrent disordering of the L$1_0$ structure begins at T $\sim$ 600 K after approximately 1 hour. No clear sign of precipitates or of L$1_0$ superlattice reflections were detectable  at 923 K and after cooling the specimen to room temperature (Figure S7). Overall, these results are consistent with the existence of a transitional phase with a magnetic transition temperature of 740 K that bridges the chemically ordered L$1_0$ FeNi phase and the disordered A1-type FeNi phase \cite{lewis2016} of Curie temperature 660 K. This study also demonstrates that, upon heating, the large population of Ni-rich and Ni-poor precipitates dissolve into the tetrataenite matrix, which itself is hypothesized to undergo a chemical disordering. These changes combine to collapse the magnetocrystalline anisotropy and yield magnetically soft behavior. Finally, the large increase in saturation magnetization noted in the third and final room temperature hysteresis loop is consistent with dissolution of the A6-type AFM ultrafine phase, which was previously providing magnetic voids that reduced the matrix saturation magnetization of the meteorite. 

\begin{figure}
	\centering
	\includegraphics[width=0.6\linewidth]{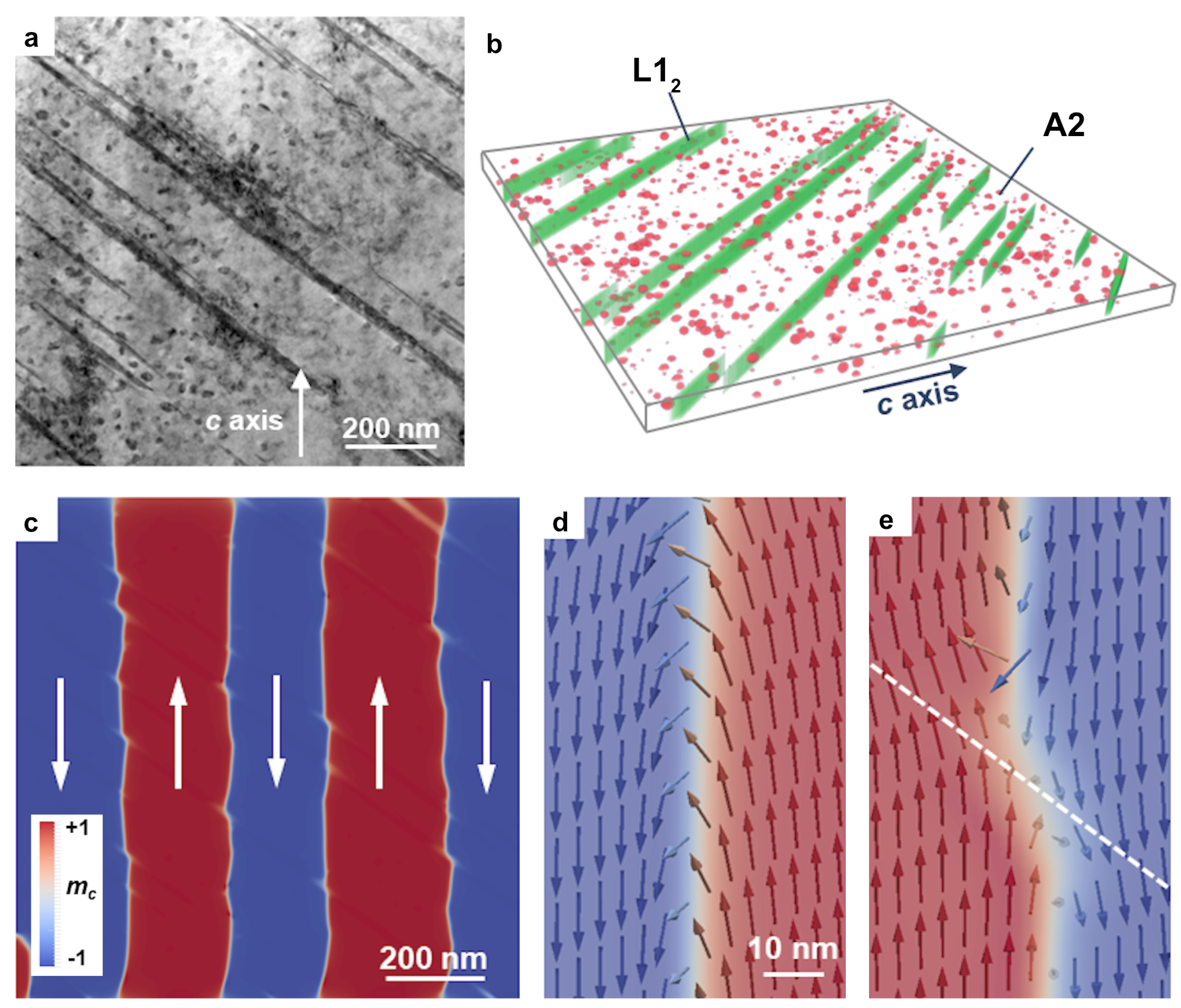}
	\caption{Micromagnetic simulations of A2, A6, and L$1_2$ Fe-Ni precipitates in an L$1_0$ FeNi matrix. (a) Simulation system based on the TEM image of the NWA 6259 specimen shown in Fig. 1e. (b) 3D visualization of the simulated structure, showing randomly-distributed spheroidal A2 and A6 Fe-Ni precipitates. (c) Simulation of an equilibrium magnetic state with domain walls parallel to the c axis of the L$1_0$ structure. (d,e) Simulations showing that a domain wall that propagates through an L$1_2$ lamella is bent. An example is marked by a white rectangle.  Magnified view showing twisting of the local magnetization at the intersection with the L$1_2$ lamella.}
	\label{fig:sim}
\end{figure}

This conceptualized micromagnetic state was simulated with a model (Figure \ref{fig:sim}) based on the microstructure derived from the imaging data of the NWA 6259 specimen (Figure \ref{fig:tem}), including the number density and type of precipitates determined from the APT experiments (Figure \ref{fig:apt}). The ferromagnetic A2-type cubic precipitates were distributed evenly and randomly throughout the sample, whereas the A6-type AFM nanoprecipitates were simulated as non-magnetic 2-nm-diameter voids with a vanishing magnetization. The resultant contour plot of the magnetization component parallel to the L$1_0$ \textit{c} axis (easy axis), Figure \ref{fig:sim}c, shows a simulated micromagnetic domain state with \ang{180} magnetic domains parallel to the \textit{c} axis with widths of approximately 200 nm that is in excellent agreement with the experiment (Figure \ref{fig:magn}b). Closer inspection of the simulated domain walls reveals a straight wall structure (Figure \ref{fig:sim}d) and distortions, or kinks, at intersections with L$1_2$ lamellae (Figure \ref{fig:sim}e), exactly as found in the electron microscopy experiment (Figure \ref{fig:magn}b). A detailed view of the local magnetization rotation and Bloch domain wall broadening at a phase intersection is shown in Figure \ref{fig:sim}e. These kinks are attributed to the difference in magnetic anisotropy, and consequently in the magnetic domain wall energy and width, between the L$1_0$ and L$1_2$ phases. The magnetic domain wall width is calculated as 5.6 nm in the L$1_0$ phase and 18 nm in the L$1_2$ phase (Fig. S5). Further, Figure \ref{fig:sim}e shows how the handedness of the domain wall differs on either side of the L$1_2$ lamella, a feature associated with minimization of the dipole-dipole energy state of the domain wall that has implications for the stability of the ensemble magnetic state. These domain walls signal magnetically weak spots where magnetization curling instabilities can form and facilitate magnetization reversal.

\section*{Scientific and technological implications of the meteoritic hidden microstructure}

New knowledge of the previously undescribed ``hidden'' structure and properties of the NWA 6259 meteorite reported here impacts not only how iron meteoritic data might be used interpret the origins of our solar system but also invites renewed consideration of tetrataenite as sustainable permanent magnet. Utilizing highest-resolution probes combined with magnetometry and simulations, the microstructure is revealed to be comprised of a magnetic phase assemblage of ferromagnetic cubic ($\sim$30 nm diameter), antiferromagnetic tetragonal ($\sim$2 nm diameter) precipitates and ferromagnetic L$1_2$-type lamellae embedded in a tetrataenite matrix. At the current time these antiferromagnetic precipitates are not considered to be the hypothesized \textit{antitaenite} phase \cite{Rancourt1995,Rancourt1997,Rancourt1999}, on the basis of different postulated formation modes, crystal structures and magnetic transition temperatures.
These antiferromagnetic precipitates decrease the saturation magnetization and the soft magnetic inclusions act as weak regions that nucleate easy magnetization reversal. Both of these effects decrease the technical magnetic properties of the meteoritic sample, advancing the deduction that the maximum energy product of tetrataenite, deduced from measurements of natural materials such as the NWA 6259 meteorite, may be underestimated by as much as 15-20 \% to reach a value that is close to 70 \% of that of the best rare-earth magnets. This conclusion invites renewed consideration of tetrataenite as a sustainable advanced permanent magnet.

\begin{acknowledgement}

The authors are grateful to D. Meertens, M. Kruth and W. Pieper for TEM specimen preparation and to M. Keil for EBSD preparation. This project has received funding from the European Research Council (ERC) under the European Union's Horizon 2020 research and innovation programme (Grant No. 856538, project ``3D MAGiC'') and from the Horizon 2020 Research and Innovation Programme (Grant No. 823717, project ``ESTEEM3''). Funding by the Deutsche Forschungsgemeinschaft (DFG, German Research Foundation) via CRC/TRR 270 ``HoMMage'' (Project-ID 405553726) is gratefully acknowledged. Portions of this research were conducted with high performance computational resources provided by the Louisiana Optical Network Infrastructure (http://www.loni.org). This research was supported in part by a cooperative agreement with U.S. Department of Energy's Advanced Research Projects Agency-Energy (ARPA-E) and by Northeastern University.

\end{acknowledgement}

\begin{suppinfo}

\begin{itemize}
  \item Filename: Methods, Materials, Figures
\end{itemize}

\end{suppinfo}


\bibliography{Tt_bibliography}

\end{document}